\documentclass[aps,prd,superscriptaddress,nofootinbib,twocolumn,10pt]{revtex4}
\usepackage{color,amsmath,amssymb,graphicx,latexsym,lineno}
\usepackage{threeparttable,multirow,txfonts,subfigure,booktabs}

\graphicspath{{./}{fig_simu/}}

\begin{document}

\title{Simulation study of the cosmic ray Sun shadow with a time-dependent solar magnetic field model}

\author{Jie Xia}
\affiliation{Division of Dark Matter and Space Astronomy, Purple Mountain Observatory, Chinese Academy of Sciences, Nanjing 210023, P. R. China}
\affiliation{School of Astronomy and Space Science, University of Science and Technology of China, Hefei 230026, P. R. China}

\author{Ming-Yang Cui}
\email[Corresponding author: ]{mycui@pmo.ac.cn}
\affiliation{Division of Dark Matter and Space Astronomy, Purple Mountain Observatory, Chinese Academy of Sciences, Nanjing 210023, P. R. China}

\author{Qiang Yuan}
\email[Corresponding author: ]{yuanq@pmo.ac.cn}
\affiliation{Division of Dark Matter and Space Astronomy, Purple Mountain Observatory, Chinese Academy of Sciences, Nanjing 210023, P. R. China}
\affiliation{School of Astronomy and Space Science, University of Science and Technology of China, Hefei 230026, P. R. China}

\author{Yi Zhang}
\affiliation{Division of Dark Matter and Space Astronomy, Purple Mountain Observatory, Chinese Academy of Sciences, Nanjing 210023, P. R. China}
\affiliation{School of Astronomy and Space Science, University of Science and Technology of China, Hefei 230026, P. R. China}

\author{Guang-Lu Shi}
\affiliation{Division of Dark Matter and Space Astronomy, Purple Mountain Observatory, Chinese Academy of Sciences, Nanjing 210023, P. R. China}

\author{Li Feng}
\affiliation{Division of Dark Matter and Space Astronomy, Purple Mountain Observatory, Chinese Academy of Sciences, Nanjing 210023, P. R. China}
\affiliation{School of Astronomy and Space Science, University of Science and Technology of China, Hefei 230026, P. R. China}

\begin{abstract}
During the propagation of cosmic rays in the solar system, the Sun will block those particles
and form a shadow whose position and depth are very important probe of the magnetic fields in the
Sun's corona, in the interplanetary space, and the Earth's vicinity. In this work we carry out
Monte Carlo studies of the Sun shadow, with a novel approach to take into account daily variations 
of the coronal and interplanetary magnetic field models. This treatment is suitable for studies of
short-term variations of the Sun shadow, which become detectable by the Large High Altitude Air Shower
Observatory (LHAASO) experiment. Two different coronal magnetic field models, the Potential Field 
Source Surface (PFSS) and Current Sheet Source Surface (CSSS) models, with observational time-varying
photospheric magnetic fields as boundary conditions, are studied in this work. The interplanetary 
magnetic fields are then derived using the Parker spiral model based on the coronal ones. Furthermore, 
both the coronal and interplanetary magnetic field strengths are corrected using the Parker Solar 
Probe (PSP) measurements. We compare the simulation results with the daily observations of Sun shadow by 
LHAASO in 2021, and find that the CSSS model generally shows better consistency of the displacement 
of the Sun shadow than the PFSS model.

\end{abstract}

\keywords{Cosmic rays --- Sun's shadow --- solar magnetic field}

\maketitle

\section{Introduction} \label{sec:intro}

The solar magnetic field is one of the most important quantities in governing a number of
solar activities. Observations of the solar magnetic fields mainly focus on the photosphere, 
where accurate measurements are feasible \cite{2021JSWSC..11....4P}. Key instruments or 
projects include the Helioseismic and Magnetic Imager (HMI; \cite{2012SoPh..275..229S}) 
and the Global Oscillation Network 
Group\footnote{https://nso.edu/telescopes/nisp/gong/} (GONG). High-resolution global and 
time-dependent magnetic maps are available \cite{2017ApJ...848...70L}. However, to get the
coronal magnetic fields (CMF), analytical or numerical models are required, with boundary
conditions of the photospheric magnetic field data. The most commonly used models are the
Potential Field Source Surface (PFSS, \cite{1969SoPh....6..442S,1969SoPh....9..131A}) and 
Current Sheet Source Surface (CSSS, \cite{1995JGR...100...19Z,2002AdSpR..29..411Z}) models. 
These models can effectively describe the steady state CMF structures at relatively large
scales \cite{2003SoPh..212..165S, Shi2024}. The CMF will be carried by the solar wind, which 
propagates in the heliosphere and forms the interplanetary magnetic fields 
(IMF; \cite{1958ApJ...128..664P}). The IMF strength and structure depend on the solar wind 
speed, the Sun's magnetic polarity, as well as solar activities, resulting in a complex 
magnetic environment within the heliosphere \cite{Priest_2014}. The Parker's spiral model 
is widely employed to describe the IMF when the Sun is in the quiet phase
\cite{1958PhRv..110.1445P,1958ApJ...128..664P}. The measurements of the IMF, such as those 
from the OMNI dataset \cite{2005JGRA..110.2104K}, are typically taken near the Earth. 
Recently, the Parker Solar Probe (PSP; \cite{2023SSRv..219....8R}) is able to measure the 
magnetic fields in the vicinity of the Sun, providing very important constraints on the solar 
magnetic field models. It was found that the model predicted IMF based on the photospheric 
magnetic fields were often underestimated compared with the measurements near the Earth
\cite{1978SoPh...58..225S,2019SoPh..294...19W}.

When Galactic cosmic rays (GCRs) propagate in the solar system, part of them will be blocked 
by the Sun, forming a shadow of the GCR intensity. Due to the deflection of charged particles
in the heliospheric magnetic field, the direction of the Sun shadow will shift from the Sun's
position and the shape also changes. As a result, observing the Sun shadow of GCRs provides a
unique tool to probe the solar magnetic fields, which is an important complement of the
direct measurements \citep{2013PhRvL.111a1101A,2018ApJ...860...13A,2024arXiv241009064T}.
Very interestingly, it has been shown that the north-south displacement of the Sun shadow
has a very good correlation with the transverse component of the IMF but with a 3.3-day time
shift, with the Sun shadow response being earlier than the in situ measurement near the Earth
\cite{2024arXiv241009064T}. This gives a possibility to forecast the magnetic field changes
using GCR observations. 

The Sun shadow has been studied by several groundbased air shower experiments. The Tibet 
AS$\gamma$ experiment studied the displacement and deficit of the Sun shadow for different 
stages of solar activities, and found that the CSSS model was favored by the observations
\cite{1993ApJ...415L.147A,2000ApJ...541.1051A,2006AdSpR..38..936A,2013PhRvL.111a1101A,2018PhRvL.120c1101A}.
The ARGO-YBJ experiment highlighted the link between the IMF sector structures and the Sun 
shadow displacement \cite{2011ApJ...729..113A,2017ICRC...35...41C}. The IceCube experiment
also observed the Sun shadow, but did not have a strong preference of the CMF models 
\cite{2019ICRC...36..437T,2021PhRvD.103d2005A}. The relative intensities of the Sun shadow 
at a cadence of $\sim27.3$ days, corresponding to one Carrington rotation (CR), were studied 
by HAWC, and the potential correlation with the photospheric magnetic fields was investigated
in detail \cite{2024ApJ...966...67A}. With the large effective area and data statistics, 
LHAASO is able to observe the daily Sun shadow with high precision \cite{2024arXiv241009064T},
which enables careful studies of the time-dependent CMF and IMF using GCRs. However, when 
comparing with the simulation predictions, usually a static magnetic field model (during 
the time window of each Sun shadow measurement) is assumed. This could be enough if the 
observation time window is relatively long (e.g., longer than one month). Since the typical 
speed of solar wind is about 500 km s$^{-1}$, which takes about $3\sim4$ days to propagate 
from the Sun to the Earth. In this sense, it is necessary to take into account the time 
evolution of the magnetic field if one wants to study the Sun shadow with a high time 
resolution. In this work, we develop a time-dependent magnetic field model to simulate 
the Sun shadow. It is expected to be very useful in detailed studies of the Sun shadow 
and solar magnetic fields, especially when there are strong solar activities. 

This paper is organized as follows. In Sec. \ref{sec:model}, we introduce the time-dependent
magnetic field models. In Sec. \ref{sec:simu} we describe the simulation of the Sun shadow. 
In Sec. \ref{sec:result} we present the simulation results and compare them with recent LHAASO
daily observations. We summarize our study in Sec. \ref{sec:sum}.

\section{Time-dependent magnetic field models} \label{sec:model}

The magnetic fields relevant to the formation of the GCR Sun shadow include three parts, 
the CMF within a few solar radii, the IMF between the Sun and the Earth, and the geomagnetic 
field (GMF) near the Earth. The solar wind carries the coronal open magnetic field, forming a 
spiral IMF that propagates outwards. When getting close to the Earth, the GMF 
becomes more and more important, and we add the model of GMF to the model of IMF to get the combined magnetic fields.
In this work, the CMF and IMF are calculated based on particular models, and the GMF is 
described by the International Geomagnetic Reference Field (IGRF-13;
\cite{2021EP&S...73...49A}).

The PFSS and CSSS models are widely used to describe the CMF in the community. The PFSS model assumes 
that the current is free in the corona, and the magnetic field distribution is derived via solving 
the Laplace equation. In contrast, the CSSS model considers additionally the horizontal current 
sheet to account for non-potential effects, enabling the simulation of complex magnetic topologies
\cite{1995JGR...100...19Z,2002AdSpR..29..411Z}. The boundary condition, specifically the photospheric
magnetic field is obtained from observations by the HMI. To accurately model the temporal variations 
of the CMF, we use the daily synoptic maps. 
For the PFSS model, the finite difference method is employed to directly solve the
Laplace's equation, with the heliocentric distance of source surface $R_{ss}=2.5~R_{\odot}$ where
$R_{\odot}$ is the Sun's radius. 
For the CSSS model, spherical harmonic expansions have been adopted to calculate the magnetic field. 
The heliocentric distance of the source surface is set to be $R_{ss}=10~R_{\odot}$. 
Besides the source surface, the heliocentric distance of the cusp surface ($R_{cs}$) is introduced,
which is set to be 1.7~$R_{\odot}$ \cite{1992SSRv...61..393K}. In this study, the expansion order 
for the CSSS model is set to be 10 \cite{2013PhRvL.111a1101A}, with the spherical harmonic coefficients
being calculated from the photospheric magnetic field for the interval of [$R_{\odot}$, $R_{cs}$] and
from the radial component of the CMF at $R_{cs}$ for the interval of ($R_{cs}$, $R_{ss}$], respectively.
In the intermediate region between $R_{cs}$ and $R_{ss}$, the CMF gradually orients towards the 
radial direction. In the outer region above $R_{ss}$, the solar wind plasma carries the magnetic 
field, and thus the magnetic field above the source surface satisfies the Parker spiral lines
\citep{1995JGR...100...19Z}. We use the python package 
{\tt pfsspy}\footnote{\url{https://zenodo.org/records/1472183}}
\cite{2020JOSS....5.2732S} and the IDL-based tool 
{\tt hccsss}\footnote{\url{http://sun.stanford.edu/~xuepu/DATA/hccsss/}} 
\cite{1995JGR...100...19Z} to calculate the CMF.

\begin{figure*}[!htbp]
\centering
\includegraphics[width=0.48\textwidth]{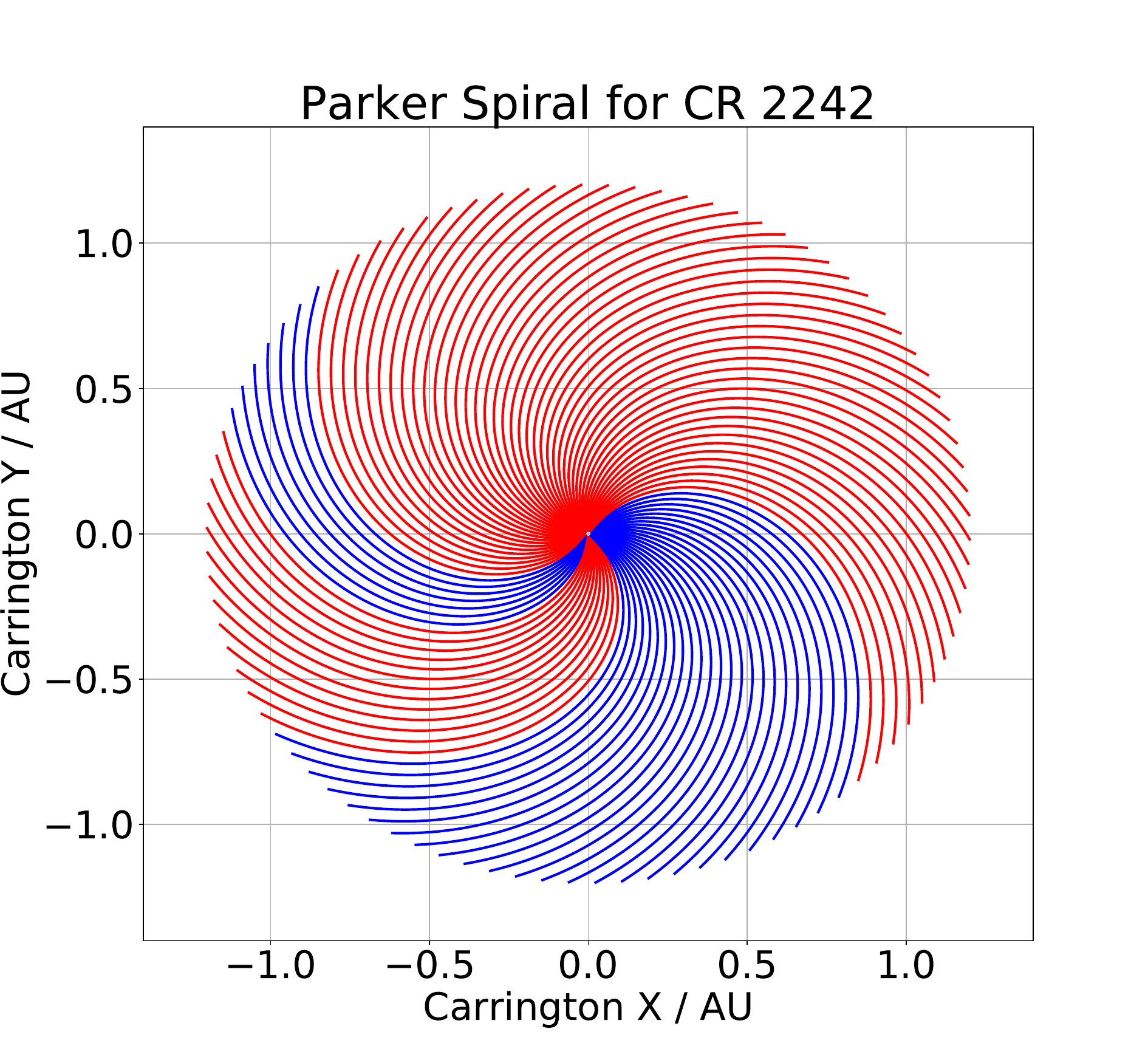}
\includegraphics[width=0.48\textwidth]{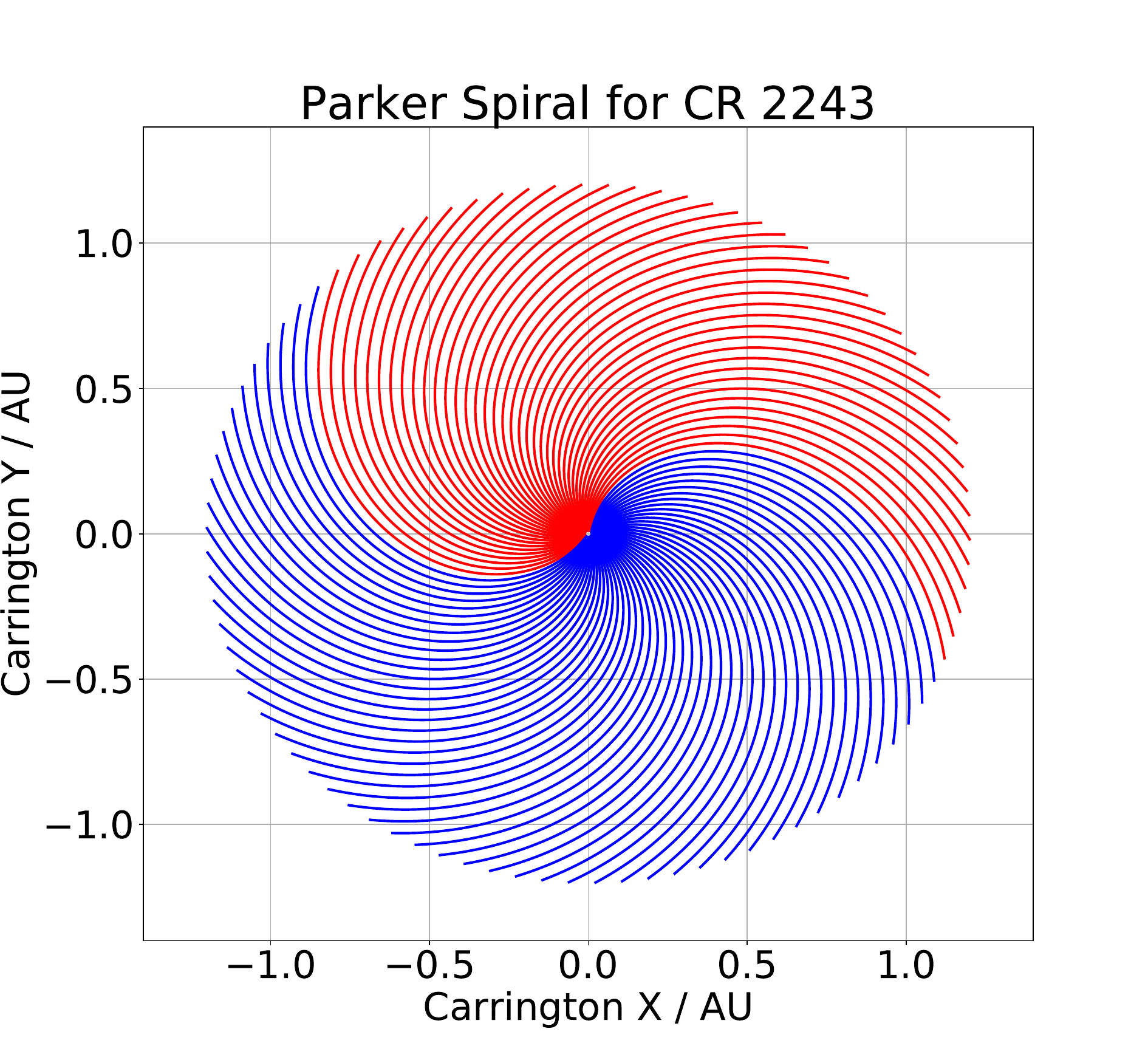}
\caption{Magnetic sectors in the solar equatorial plane, for CR 2242 (starting at 
22:36 UT, 2021-03-17; left panel) which shows 4 sectors and CR 2243 (starting at 05:32 UT, 2021-04-14;
right panel) which shows 2 sectors. Red and blue color shows positive and negative magnetic
polarities. Each field line is traced back to the outer boundary at $2.5R_{\odot}$ according 
to the Parker spiral model, assuming the PFSS model.}
\label{fig:ParkerLine}
\end{figure*}

The IMF is obtained from the computed CMF at $R_{ss}$, following the Parker spiral model
\textcolor{blue}{ \citep{1958ApJ...128..664P} }. Specifically, for $r>R_{ss}$, the IMF can be described 
by the following equations:
\begin{eqnarray}
B_r(r,\theta,\phi,t) &=& B_r\left(R_{ss},\theta,\phi',t'\right) \left(\frac{R_{ss}}{r}\right)^2, \\
B_\theta(r,\theta,\phi,t) &=& 0, \\
B_\phi(r,\theta,\phi,t) &=& {-} B_r(r,\theta,\phi,t) \left(\frac{\Omega_{\odot}}{V_{\rm sw}(\theta)}\right) (r - R_{ss}) \sin \theta,
\end{eqnarray}
where $V_{\rm sw}$ is the solar wind speed, $\Omega_{\odot}$ is the rotational angular speed of the 
Sun. For $r<R_{ss}$, the CMF models are used. The longitude and time in the interplanetary
space are related to those at the source surface as follows. Here, $t$ denotes the
observation time at the point $(r,\theta,\phi)$ in the interplanetary space, and $t'$ represents the
corresponding time at the source surface ($R_{ss}$) when the solar wind (which carries magnetic field 
to the point) was emitted from the source surface. Variables $\phi'$ and $t'$ can be obtained as
\begin{eqnarray}
\phi' &=& \phi + \frac{R_{ss}  \Omega_{\odot}}{V_{\rm sw}(\theta)} \left[\frac{r}{R_{ss}} -1 -\ln\left(\frac{r}{R_{ss}}\right)\right], \\
t' &=& t-(r-R_{ss})/V_{\rm sw}(\theta).
\end{eqnarray}
These relations reflect the fact that magnetic fields propagate from the source surface outwards
following spiral trajectories along with the solar winds. As an illustration, the topology of the 
Parker spiral for two CRs is shown in Fig.~\ref{fig:ParkerLine}, plotted in the
solar equatorial plane ($0^\circ$ latitude). The maps correspond to CRs 2242 and 2243, respectively, 
assuming the PFSS model of the CMF. Different color represents different polarities of the magnetic 
field. We can see that the magnetic field structures are complicated and varying with time. For CR 
2242 there are 4 alternating sectors of polarities, and for CR 2243 there are only 2 sectors. 
Even more complicated case with 6 sectors also exist.  

The solar wind speed is important in determining the IMF from the CMF. 
The solar wind is highly complex, with an average speed of approximately 400 km~s$^{-1}$, 
and the maximum speed can reach thousands of kilometers per second.
The solar wind is directly modulated by solar activities such as CMEs, which have an average
rate of $0.2$ day$^{-1}$ during the solar minimum and increase to $\sim 3$ day$^{-1}$ during 
the solar maximum \cite{1958PhRv..110.1445P}. In this work we use the speed measured by the 
interplanetary scintillation \cite{2022ApJS..259....2P}. The variation of solar wind speed with 
latitudes is taken into account. During the solar minimum, the solar wind speed in high-latitude 
regions is higher than that in low-latitude regions, while during the solar maximum, the 
difference between high and low latitudes tends to disappear \cite{2022ApJS..259....2P}.
This variation law is well consistent with the periodic evolution of sunspot numbers. 
Regarding the radial distribution, we assume the solar wind speed to be constant along the radial direction for a given latitude.
The detailed latitude dependence of the solar wind speed in 2020 is given in Fig.~\ref{fig:SW} 
in Appendix \ref{sec:SW}. For years 2021-2023, we directly use the 2020 results as an
approximation.

\begin{figure}[!htbp]
\centering
\includegraphics[width=0.48\textwidth]{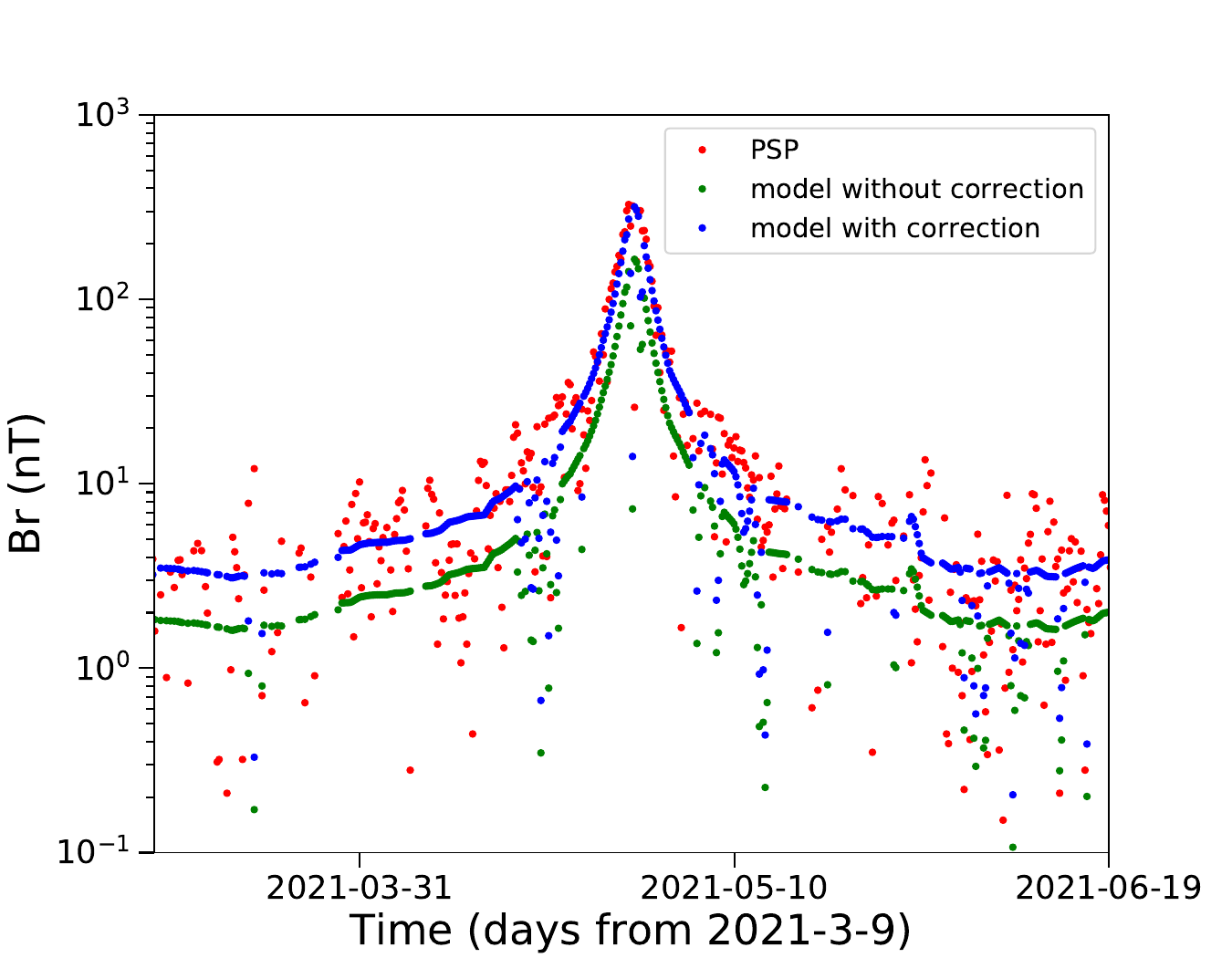}
\caption{Comparison of $B_r$ as measured by PSP (red dots) and predicted by the CSSS model
(green dots). Blue dots show the corrected model predictions.}
\label{fig:CorrPSP}
\end{figure}

It is important to note that CMF models tend to systematically underestimate the actual field 
strengths and some corrections are required \cite{1978SoPh...58..225S,1995ApJ...447L.143W,2017ApJ...848...70L,2020ApJS..246...23B}. 
The PSP observations close to the Sun are very useful in calibrating the model calculation.
We perform the magnetic field correction for each orbit of the PSP.
The correction factor is determined through minimizing the $\chi^2$ statistic between PSP 
observations and theoretical model predictions over the period of one orbit. As an illustration, 
the comparison between the PSP measurements (red dots) of the radial component $B_r$ and the 
CSSS model predictions (green dots) from March to June of 2021 is shown in Fig.~\ref{fig:CorrPSP}. 
We find that via up-scaling the model results by a factor of $\sim1.92$, the corrected results 
(blue dots) are in better agreement with the observations. The correction factors of the PFSS 
and CSSS models for different time window are given in Table \ref{tab:correction_factor}, and 
the corrected magnetic fields $B_r$ for all PSP orbits from 2021 to 2023 are given in Appendix
\ref{sec:psp_and_corr}. 

\begin{table*}[htbp]
\centering
\caption{Correction factors of the magnetic field strength $B_r$ for the PFSS and CSSS models.}
\begin{tabular}{cccccccccccc}
\hline\hline
Time Window & {1} & {2} & {3} & {4} & {5} & {6} & {7} & {8} & {9} & {10} & {11} \\
\hline
PFSS & 10.02 & 11.94 & 4.29 & 4.84 & 2.55 & 1.95 & 2.95 & 3.49 & 1.39 & 2.06 & 1.67 \\
CSSS & 1.74 & 1.92 & 1.54 & 2.01 & 1.43 & 1.35 & 1.71 & 1.31 & 1.32 & 1.19 & 1.00\\

\hline\hline
\end{tabular}
\label{tab:correction_factor}
\end{table*}

\section{Monte Carlo Simulation} \label{sec:simu}

In this study, we adopt the ``backtracking method''  to simulate the particle propagation. 
The working coordinate system is the geographic coordinate. The CMF and IMF 
are originally computed in the Carrington heliographic coordinate system, which are then converted 
to the geographic coordinate. The position of the Sun is initially determined in the equatorial
coordinate system, which is also transformed to the geographic coordinate when performing the particle trajectory simulation.
We use the astropy package \citep{2013A&A...558A..33A} to transform the coordinates between the different reference systems.
Negatively charged particles within $4^{\circ}$ of the
Sun's center are emitted.
The paths of these particles will get deflected by the magnetic fields. We employ the fourth-order
Runge-Kutta algorithm to track particle motion in the magnetic fields. A particle hitting the Sun 
is marked as a ``missing event'', meaning that its antiparticle can not hit the detector following 
an inverse trajectory. Via collecting these missing events, we obtain the simulated Sun shadow.

In this work, we try to compare the simulation results with the recent LHAASO observations
\cite{2024arXiv241009064T}. Therefore we set the simulation parameters to be similar with
those adopted in the data analysis. The longitude and latitude of the initial 
position of particles are set to be the same values of the LHAASO experiment ($100.01^{\circ}$E,
$29.35^{\circ}$N), and the height is set as the characteristic height of the 
first interaction for particles with energies interested in this work which is $\sim20$ km.
With the movement of the Sun in the field-of-view, the zenith and azimuth angles of the Sun
vary with time. The detector's efficiency depends on the zenith angle. We thus weight the
number of simulated events to get a zenith angle distribution consistent with that observed 
by LHAASO and a uniform distribution of the azimuth angles.
The energy distribution of simulated particles is assumed to be a log-normal form, with mean 
value of $\log(E/{\rm TeV}) = 1.6$ and a width of $0.3$. The point spread function (PSF) of 
$\sim 0.5^{\circ}$ \cite{2024arXiv241009064T} is convolved to simulated events. As for the 
composition of incident particles, protons, helium, carbon, nitrogen, aluminum, and iron 
nuclei are considered, according to the abundance model of Ref.~\citep{2003APh....19..193H}.

The Sun shadow is characterized by its shift with respect to the actual position of the Sun,
and the deficit ratio due to the shielding of the Sun. We use the two-dimensional Gaussian 
function to fit the map of the Sun shadow
\begin{equation}
f(x, y) \propto \frac{1}{2\pi\sigma_x\sigma_y} \exp\left(-\frac{(x - \mu_x)^2}{2\sigma_x^2} - \frac{(y - \mu_y)^2}{2\sigma_y^2}\right),
\end{equation}
where $x$ represents the east-west direction and $y$ represents the north-south direction,
$\mu_x$ ($\mu_y$) and $\sigma_x$ ($\sigma_y$) are the central position and width of the shadow. 
The two-dimensional Gaussian is to describe the asymmetric morphology of the 
Sun shadow, since the north-south displacement is mainly affected by the IMF, while the east-west
displacement is jointly affected by the IMF and GMF.
The deficit ratio in the simulation is calculated as
\begin{equation}
\text{Deficit Ratio} = \frac{-N_{\text{hit}}}{N_{\text{all}}},
\end{equation}
where $N_{\rm all}$ is the total number of simulated events within $1^{\circ}$ radius region 
centered on the Sun without considering the deflection by the magnetic fields, and $N_{\rm hit}$ 
represents the number of events hitting the Sun whose original directions are within the same
region for the calculation of $N_{\rm all}$, after including the magnetic fields. Note that the 
PSF has been convolved in the simulation.

\section{Result} \label{sec:result}

\begin{figure}[!htbp]
\centering
\includegraphics[width=0.48\textwidth]{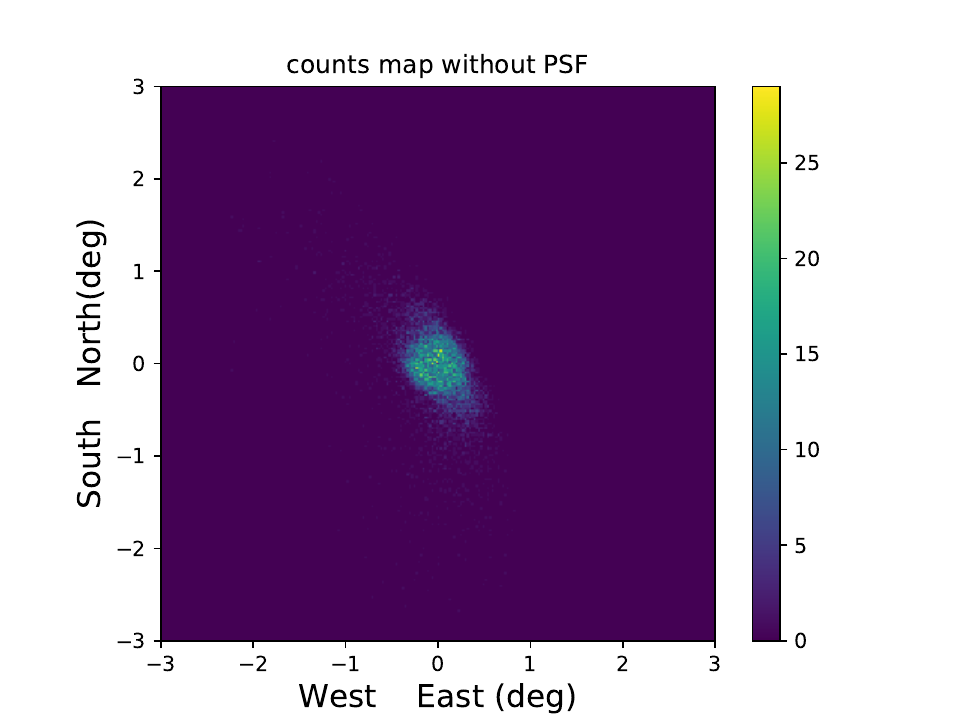}
\includegraphics[width=0.48\textwidth]{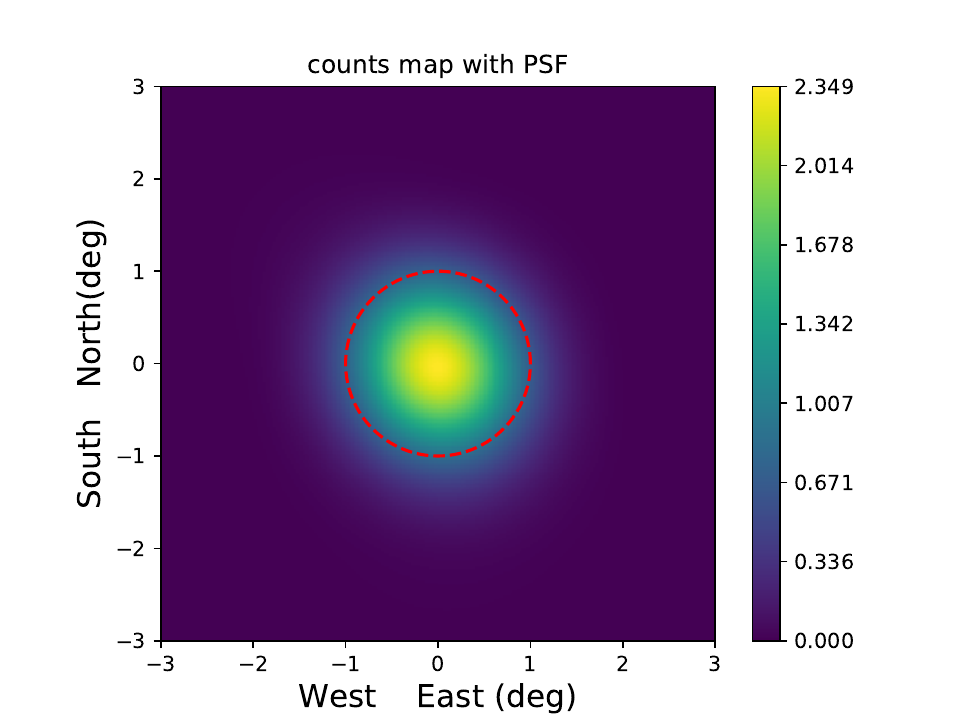}
\caption{Simulated count maps of negative charge particles in the CR 2243 for the CSSS model. 
The top panel shows the results without convolution of the PSF, and the bottom panel shows the 
results convolved with the PSF. The mean energy of the events is about 40 TeV. 
The red circle labels the $1^{\circ}$ radius region which is chosen to calculate the deficit ratio.}
\label{fig:CM}
\end{figure}

\begin{figure}[!htbp]
\centering
\includegraphics[width=0.48\textwidth]{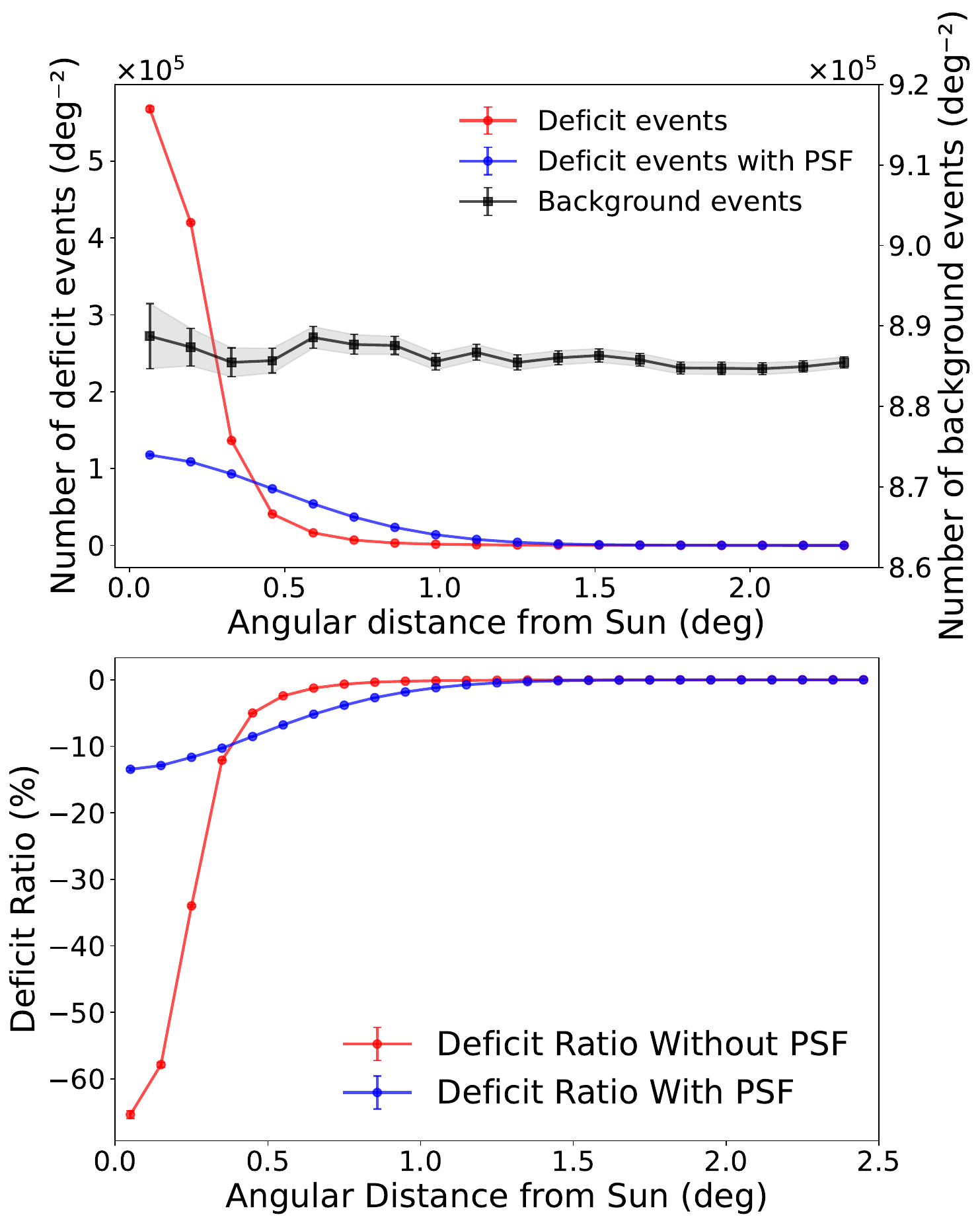}
\caption{Simulated number of deficit events and background events (top panel) and deficit ratios
(bottom panel) as functions of angular distance from the Sun. Both the results without and with 
the convolution of the PSF are presented.}
\label{fig:event_dis_ang}
\end{figure}

\begin{figure}[!htbp]
\centering
\includegraphics[width=0.48\textwidth]{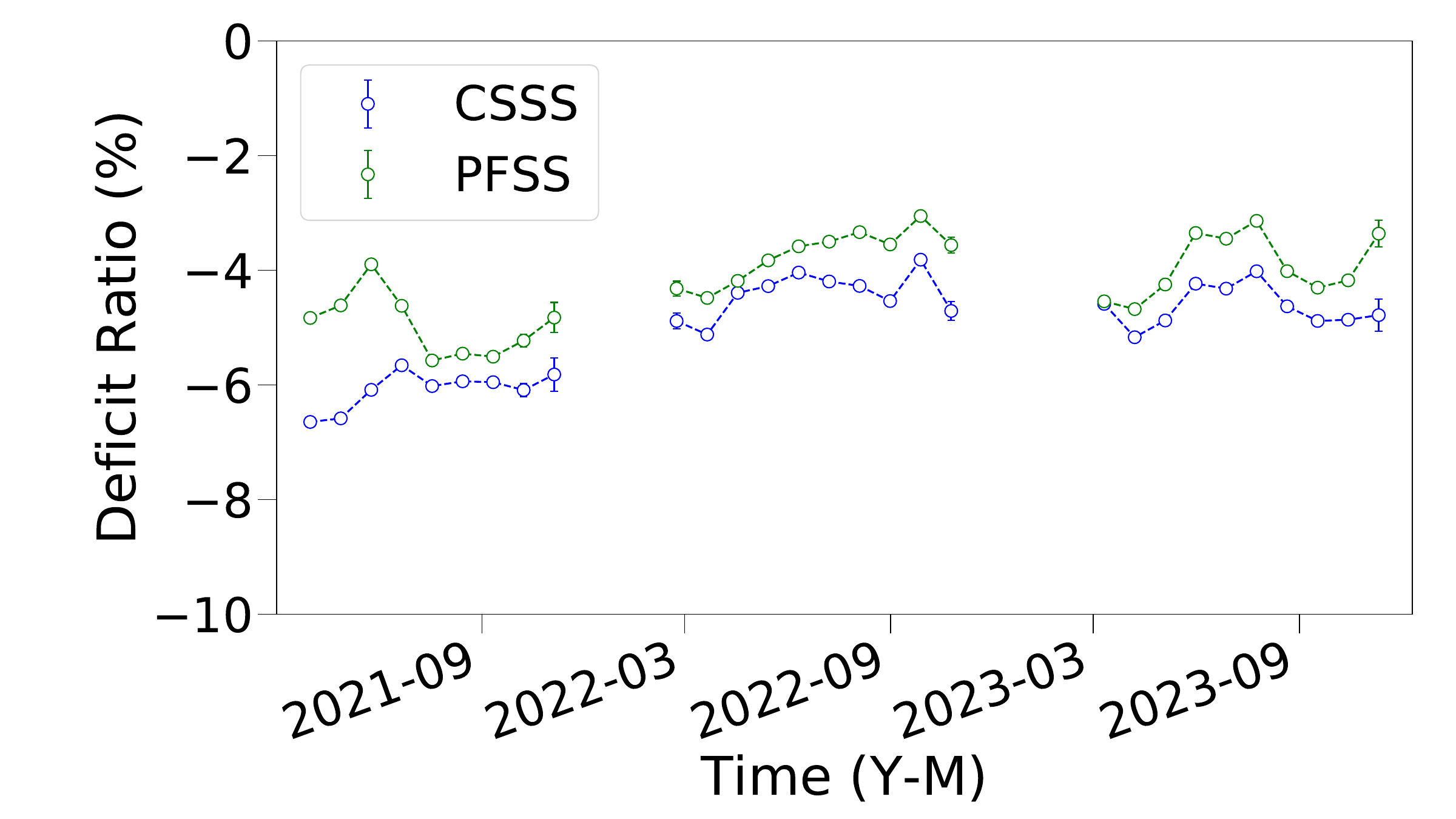}
\includegraphics[width=0.48\textwidth]{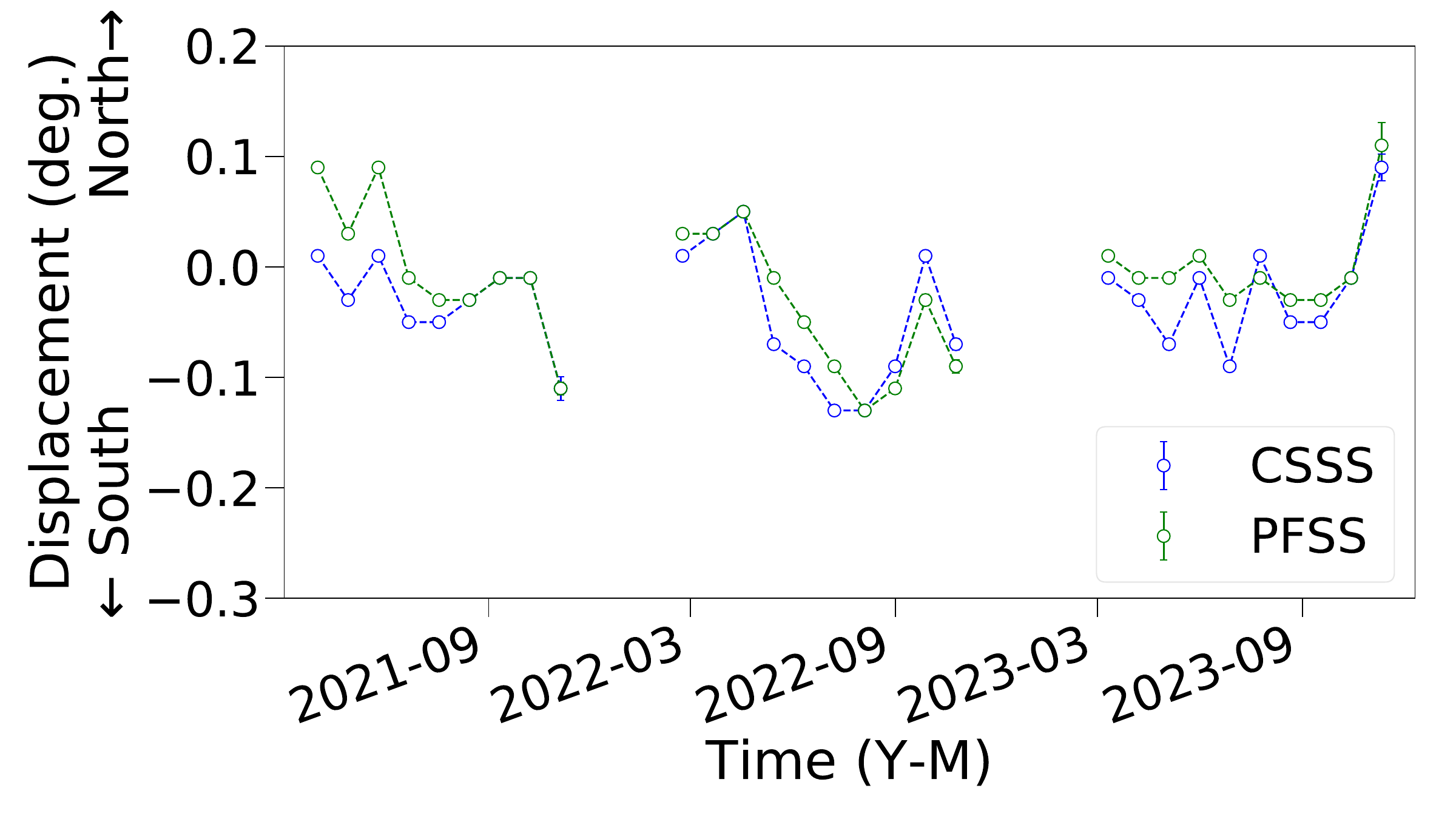}
\includegraphics[width=0.48\textwidth]{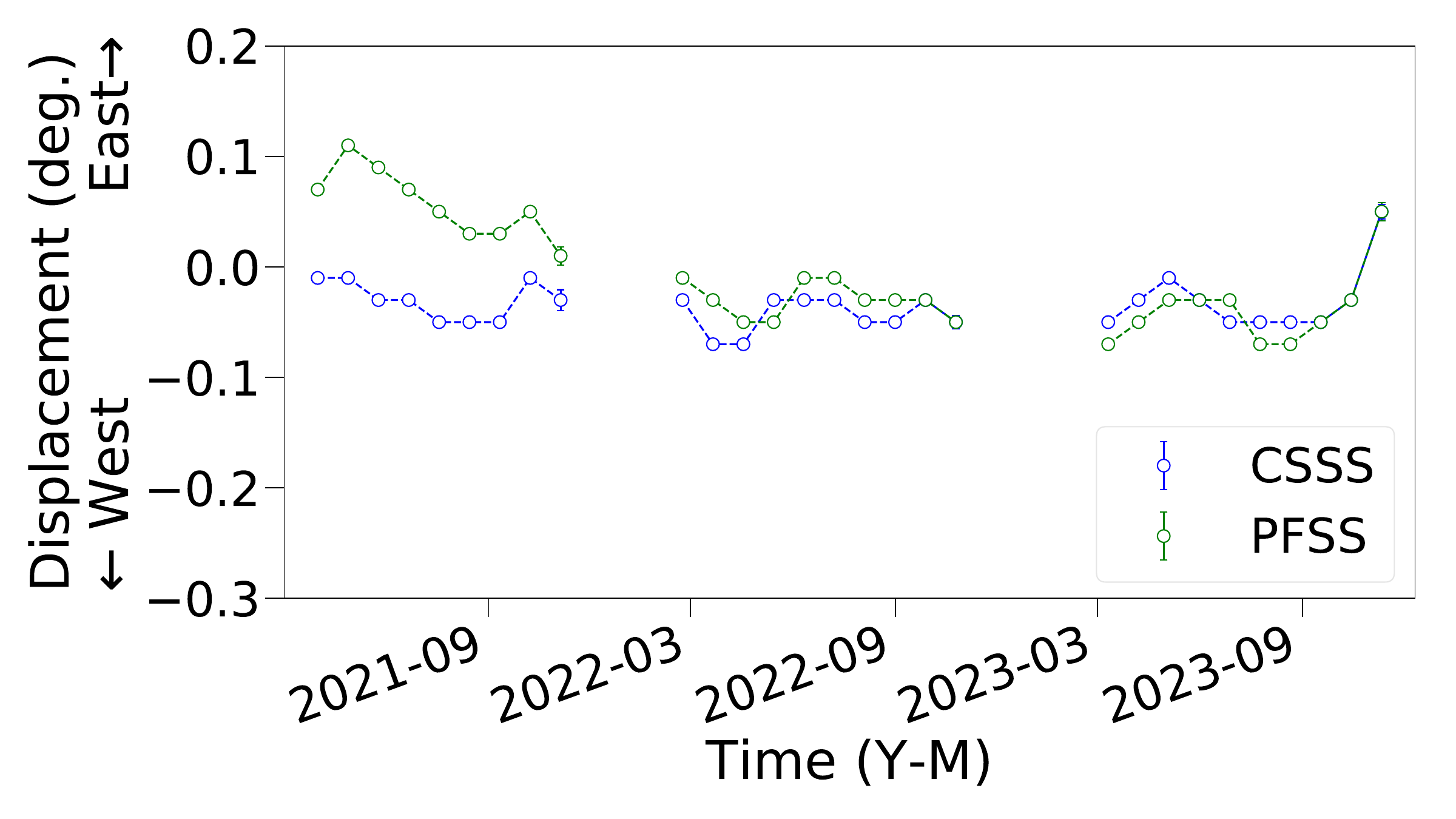}
\caption{Simulated results of deficit ratios (top), north-south (middle) and east-west displacements
(bottom) for CRs from 2242 (March, 2021) to 2277 (November, 2023). The mean energy 
of the events is about 40 TeV.}
\label{fig:DR_and_Dis}
\end{figure}

The simulated count maps of negative charge particles hitting the Sun during 
the CR 2243 are shown in Fig.~\ref{fig:CM}, with and without the PSF, respectively. The CMF model 
used is the CSSS model. To better view the angular distributions of deficit events, we show in
Fig.~\ref{fig:event_dis_ang} the one-dimensional distributions of simulated deficit events (top) 
and deficit ratios (bottom) with respect to the angular distance from the Sun. We can see that 
if the PSF is not considered, most of the deficit events are within $0.5^{\circ}$ radius region, 
which is bigger than the disk size of the Sun, showing the effect of the magnetic fields. 
The PSF effect further smooths out the shadow into bigger area.

Using Eqs. (6) and (7), we can derive the east-west and north-south displacements, as well as the 
deficit ratios. The results for CRs from 2242 to 2277, for the PFSS and CSSS models, respectively, 
are shown in Fig.~\ref{fig:DR_and_Dis}. The top panel is for the deficit ratios, and the middle 
panel is for the north-south displacements. It can be seen that with the increase of solar activities, 
the Sun shadow becomes weaker and the deficit ratio is closer to 0. Overall, the CSSS model predicts 
more deficits than the PFSS model, which can be tested with future observations by e.g., LHAASO. 
The north-south displacements show no clear trend with solar activities. However, this may be due 
to that the average of one CR washes out even bigger fluctuations of the displacements as will be 
discussed below. This emphasizes the importance of studies with higher time resolution.

\begin{figure*}[!htbp]
\begin{center}
\includegraphics[width=0.92\textwidth]{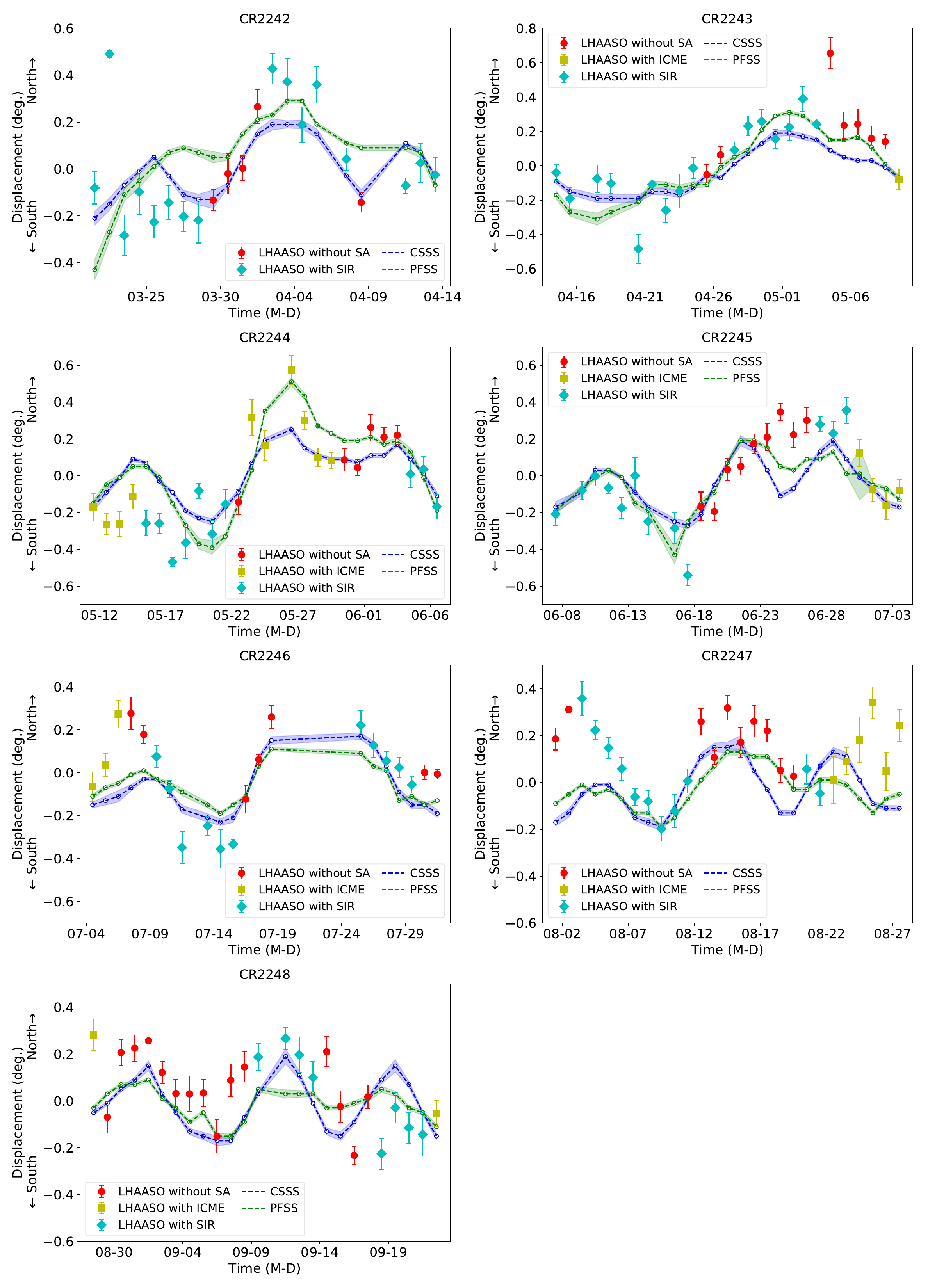}
\caption{The result of displacement with different Carrington Rotations. The panels show observed data (solid circle) and simulated results (open circle and dashed lines) for Carrington Rotations 2242 to 2248, respectively.}
\label{fig:KS}
\end{center}
\end{figure*}

\begin{table*}[htbp]
\centering
\caption{Comparison of the $\chi^2$ values over the number of data points and the corresponding 
$p$-values between simulation results and the LHAASO observations, for the CSSS, PFSS magnetic 
field models and the null hypothesis, with and without SA, respectively.}
\setlength{\tabcolsep}{10pt}  
\begin{tabular}{ccccccc}
\hline\hline
Carrington rotation & {CSSS with SA} & {CSSS w/o SA} & {PFSS with SA} & {PFSS w/o SA} & {Null with SA} & {Null w/o SA}\\
\hline
2242 & $563.54/21$       & $4.04/5$         & $562.98/21$      & $51.40/5$       & $1449.66/21$    & $34.36/5$       \\
  ($p$-value)   & $7.31\times10^{-106}$ & $5.44\times10^{-1}$  & $9.56\times10^{-106}$ & $7.17\times10^{-10}$  & $2.16\times10^{-294}$ & $2.02\times10^{-6}$  \\
\midrule

2243 & $141.55/24$       & $72.08/7$        & $98.01/24$       & $44.75/7$       & $548.31/24$     & $87.68/7$       \\
   ($p$-value)  & $1.21\times10^{-18}$ & $5.61\times10^{-13}$ & $6.54\times10^{-11}$ & $1.53\times10^{-7}$  & $1.48\times10^{-100}$ & $3.70\times10^{-16}$ \\
\midrule

2244 & $312.68/26$       & $9.39/6$         & $283.17/26$      & $14.30/6$       & $622.91/26$     & $54.20/6$       \\
  ($p$-value)  & $6.10\times10^{-51}$ & $1.53\times10^{-1}$  & $4.78\times10^{-45}$ & $2.64\times10^{-2}$  & $9.89\times10^{-115}$ & $6.74\times10^{-10}$ \\
\midrule

2245 & $205.36/25$       & $134.92/9$       & $162.19/25$      & $67.33/9$       & $358.84/25$     & $120.91/9$      \\
   ($p$-value)  & $2.84\times10^{-30}$ & $1.15\times10^{-24}$ & $4.62\times10^{-22}$ & $5.07\times10^{-11}$ & $7.79\times10^{-61}$  & $8.72\times10^{-22}$ \\
\midrule

2246 & $183.87/21$       & $105.26/7$       & $216.90/21$      & $84.92/7$       & $399.10/21$     & $66.40/7$       \\
  ($p$-value)   & $5.22\times10^{-28}$ & $8.83\times10^{-20}$ & $1.66\times10^{-34}$ & $1.36\times10^{-15}$ & $1.43\times10^{-71}$  & $7.87\times10^{-12}$ \\
\midrule

2247 & $671.65/27$       & $496.35/10$      & $806.34/27$      & $623.09/10$     & $776.51/27$     & $637.62/10$     \\
   ($p$-value)  & $3.25\times10^{-124}$& $2.66\times10^{-100}$& $1.80\times10^{-152}$& $1.98\times10^{-127}$ & $3.37\times10^{-146}$ & $1.52\times10^{-130}$\\
\midrule

2248 & $196.17/25$       & $120.56/15$      & $311.21/25$      & $236.47/15$     & $680.30/25$     & $593.62/15$     \\
  ($p$-value)   & $1.67\times10^{-28}$ & $1.47\times10^{-18}$ & $3.36\times10^{-51}$ & $7.51\times10^{-42}$ & $1.85\times10^{-127}$ & $8.05\times10^{-117}$\\
\midrule

ALL  & $2274.82/169$     & $942.60/59$      & $2440.80/169$    & $1122.27/59$     & $4835.62/169$   & $1594.78/59$    \\
  ($p$-value)   & $0.00$ & $2.09\times10^{-159}$& $0.00$ & $2.88\times10^{-196}$& $0.00$ & $1.58\times10^{-294}$\\
\hline\hline
\end{tabular}
\label{tab:fit}
\end{table*}

Fig.~\ref{fig:KS} gives the comparison of daily north-south displacements between simulations and
observations by LHAASO-KM2A \cite{2024arXiv241009064T}, from CR 2242 to 2248. Depending on whether
there were strong solar activities, the data are marked in different color for days without solar
activities (SA; red), with interplanetary coronal mass ejections (ICME \cite{2010SoPh..264..189R}; 
brown), with stream interaction regions (SIR \citep{Alvestad2021}; cyan). Compared with 
Fig.~\ref{fig:DR_and_Dis}, we see that daily displacements of the shadow are indeed much bigger
than the CR average, showing more details of the magnetic fields relevant to solar activities. 
While for some CRs the simulations can match with the observations relatively well, for the others 
they show significant deviations from each other. 

To better see the agreement between simulations and data, we calculate the $\chi^2$
statistics between the data and the model, $\chi^2=\sum_i({\rm data}_i-{\rm model}_i)^2/\sigma_i^2$, 
where $\sigma_i^2$ is the quadratic sum of the measurement uncertainty and the simulation uncertainty. 
The results are presented in Table~\ref{tab:fit}. Here the parameters of the PFSS and CSSS models
are fixed, without being optimized to better match the data.
We see that for most of CRs, the CSSS model describes the data better than the PFSS model. 
Compared with the PFSS model, the CSSS model incorporates additionally the horizontal current sheet
that can better describe the evolving CMF structures. Nevertheless, if all the data are included 
for comparison, both models show relatively large $\chi^2$ values, indicating that they are not precise 
enough to describe the real magnetic fields in corona and the interplanetary space. If we remove 
the data with ICME and SIR, the $\chi^2$ values become smaller for several CRs. This implies that 
transient events from the Sun may not be well modelled by the current model. However, for the other 
CRs the data-model matches are still poor even we do not include the data with ICME and SIR.
The results may suggest that there is limitation of the Parker spiral model for the IMF. 
As a comparison, we also give the $\chi^2$ values for the null hypothesis (i.e., 
zero magnetic field), and find that the data deviate much more severely from the null hypothesis.

Note that, there might also be uncertainties from the solar wind speed used in this work. 
As a more detailed solar wind speed model is lack, we employ a yearly average of its latitude
distribution in 2020 \cite{2022ApJS..259....2P}, and apply it to 2021-2023 which is relevant
to the current study. More detailed modeling of the solar wind speed may improve the accuracy 
of the IMF models and hence the predictions of the Sun shadow. Furthermore, future observations 
from Solar Polar-orbit Observatory (SPO; \cite{2025ChJSS..45..913D}) measuring latitudinal 
variations in solar wind velocity would provide critical data to further advance these models.

\section{Summary} \label{sec:sum}
We conduct Monte Carlo simulations of the Sun shadow of GCRs in this work. 
Compared with previous works, a major improvement is the development of the 
time-dependent solar magnetic field model which incorporates the CMF with the daily observational
photospheric magnetic field map as boundary conditions, the IMF following the Parker spiral model 
based on the CMF. As for the CMF, we consider the PFSS and CSSS models, 
which are further corrected with the PSP measurements. This time-dependent magnetic field model 
enables us to study the Sun shadow with a time resolution of one day, and is expected to be useful 
in scrutinizing the impacts on particle propagation from fast solar activities.

We compare the simulated results about the north-south displacements with the daily Sun shadow 
observations by LHAASO in 2021, and find that in general the CSSS model matches the measurements 
better than the PFSS one. Note that, the daily fluctuations of the shadow displacements are much 
larger than the average results over the CR period, highlighting the necessity of high
time-resolution studies. It has been shown that after removing the time with significant solar 
activities, the consistency between simulations and observations improves for some CRs but is
still not very good for the others. These results imply that further efforts to improve the
modeling of the magnetic field when there are transient solar events as well as in relatively
quiet period of the Sun are needed.

The data that support the findings of this article are openly
available\footnote{https://doi.org/10.57760/sciencedb.space.03566}.

\acknowledgments

This work is supported by the National Natural Science Foundation of China (Nos. 12321003 and 12220101003), 
the Project for Young Scientists in Basic Research of Chinese Academy of Sciences (Nos. YSBR-061 and YSBR-092), and the Strategic Priority Research Program of the Chinese Academy of Sciences (No. XDB0560000).

\appendix
\setcounter{figure}{0}
\renewcommand\thefigure{A\arabic{figure}}

\section{Latitude-dependence of solar wind speed} 
\label{sec:SW}

\begin{figure}[!htbp]
\centering
\includegraphics[width=0.48\textwidth]{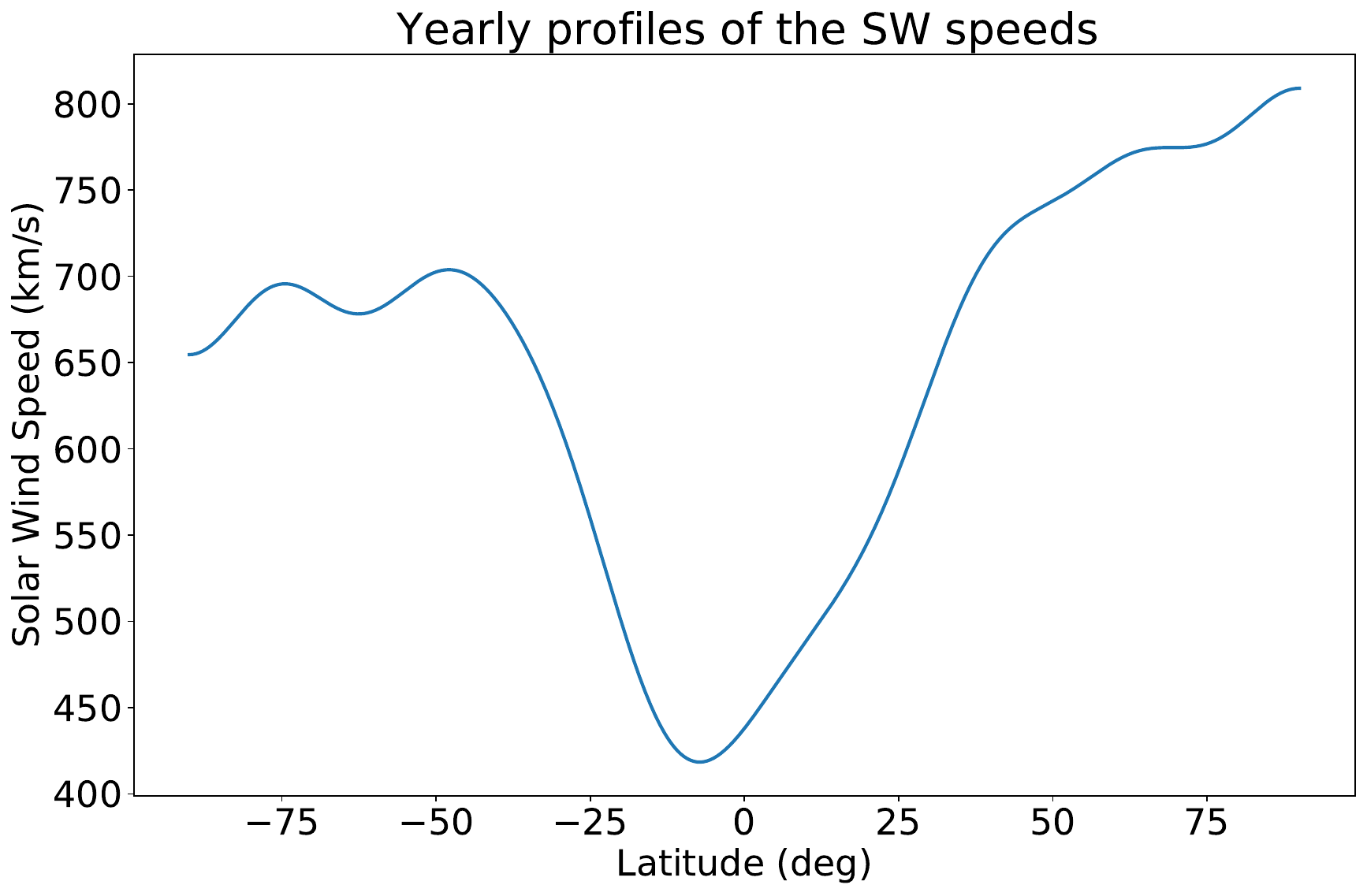}
\caption{Yearly average latitude distribution of the solar wind speed for the year 2020.}
\label{fig:SW}
\end{figure}

Fig.~\ref{fig:SW} show the latitude distribution of the solar wind speed for the year 2020
\cite{2022ApJS..259....2P}. This relation is directly employed for years 2021-2023. 

\section{Correction of $B_r$ using PSP measurements} 
\label{sec:psp_and_corr}

\begin{figure*}[!htbp]
\centering
\begin{minipage}[b]{1.0\textwidth}
  \centering
  \includegraphics[width=\textwidth]{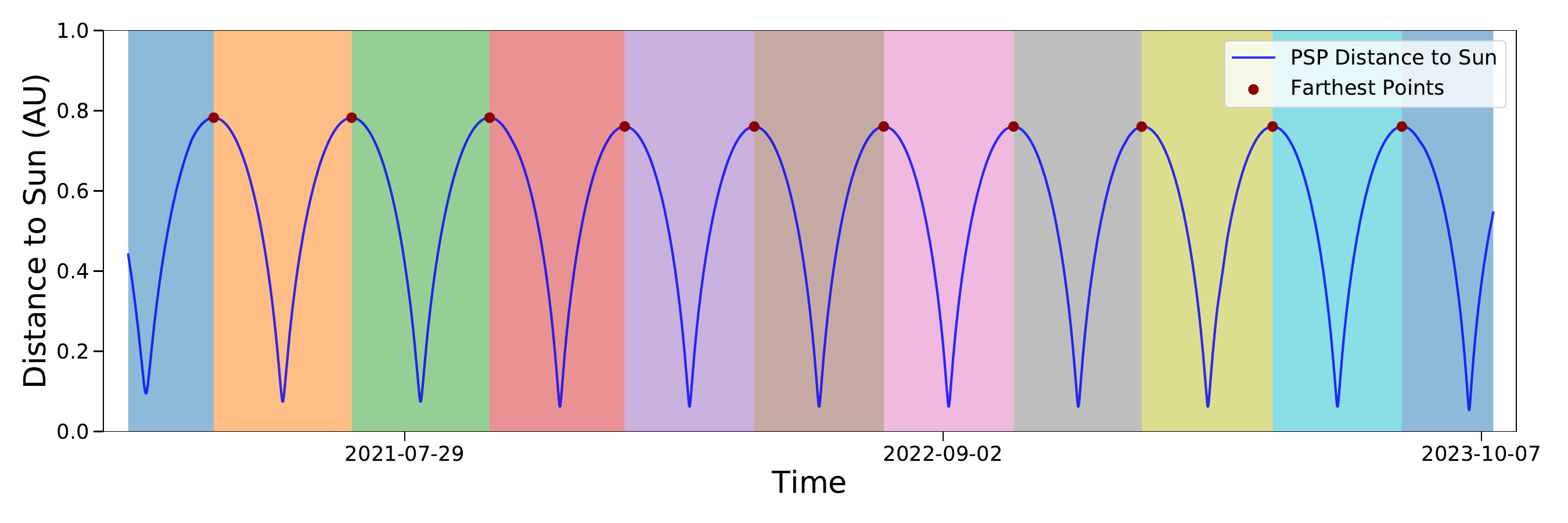}
\end{minipage}
\hfill
\begin{minipage}[b]{1.0\textwidth}
  \centering
  \includegraphics[width=\textwidth]{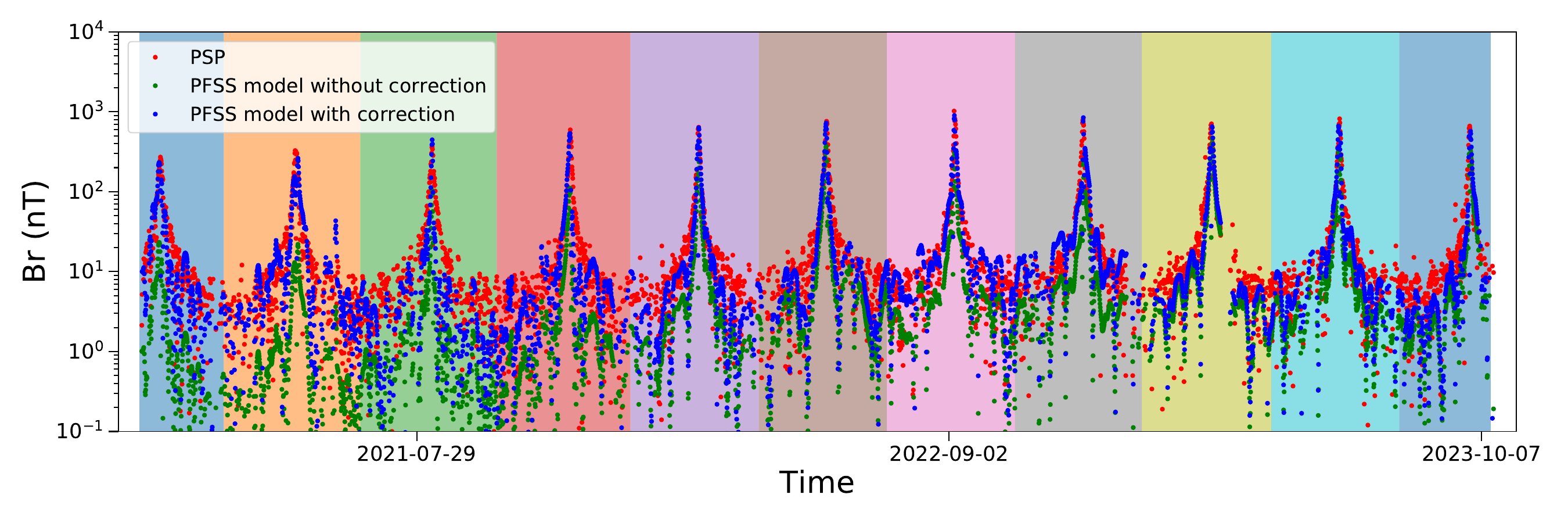}
\end{minipage}
\hfill
\begin{minipage}[b]{1.0\textwidth}
  \centering
  \includegraphics[width=\textwidth]{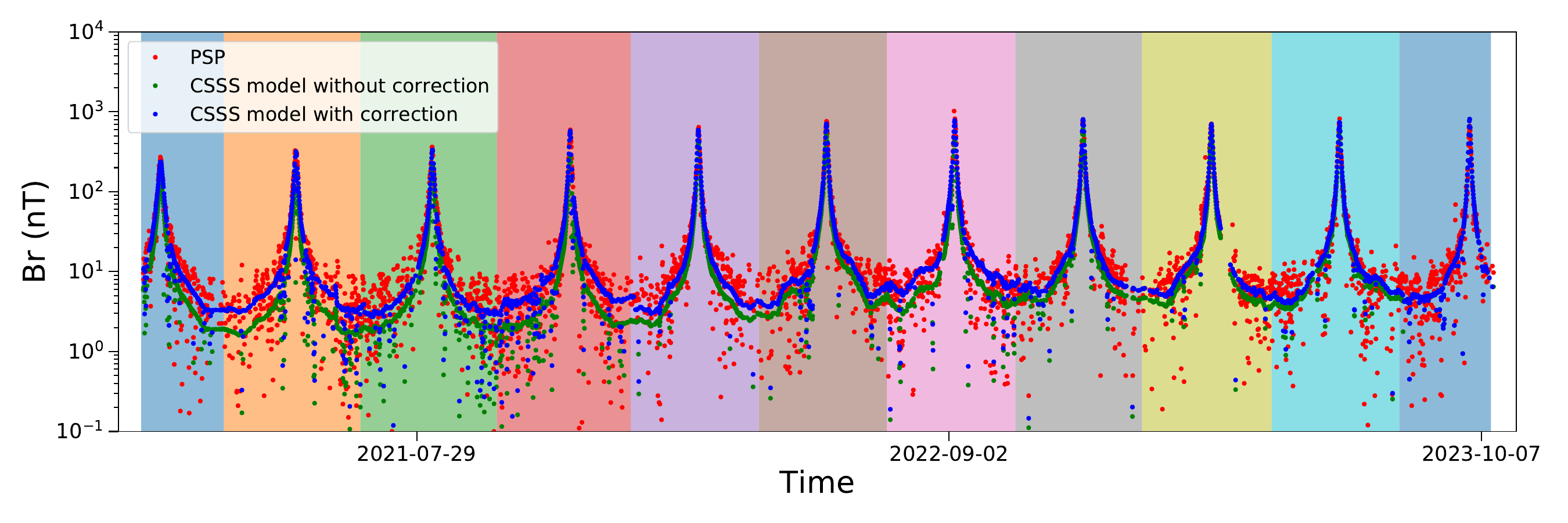}
\end{minipage}
\caption{Top: the variation of the distance between the PSP and the Sun over time. Middle: comparison between the PSSS model predicted $B_r$ (green for original prediction, blue for corrected) and PSP observations (red). Bottom: same as the middle panel but for the CSSS model.}
\label{fig:psp_and_corr}
\end{figure*}

Fig.~\ref{fig:psp_and_corr} gives the distance to the Sun for the PSP mission (top panel), the 
expected and corrected results on $B_r$ for the PFSS model (middle panel), and the expected and 
corrected results on $B_r$ for the CSSS model (bottom panel).

\bibliographystyle{apsrev}
\bibliography{refs}

\end{document}